\documentclass[12pt,preprint]{aastex}

\newcommand{\dqher}{DQ~Her}

\newcommand{\xmm}{{\sl XMM-Newton\/}}
\newcommand{\chandra}{{\sl Chandra\/}}
\newcommand{\rosat}{{\sl ROSAT\/}}
\newcommand{\euve}{{\sl EUVE\/}}
\newcommand{\einstein}{{\sl Einstein\/}}

\newcommand{\nh}{N$_{\rm H}$}

\shorttitle{The Origin of Soft X-rays in DQ Her}
\shortauthors{Mukai et al.}

\begin{document}


\title{The Origin of Soft X-rays in DQ Herculis}


\author{K. Mukai\altaffilmark{1,2}, M. Still\altaffilmark{2}}
\affil{Code 662, NASA/Goddard Space Flight Center, Greenbelt, MD 20771, USA.}

\and

\author{F. A. Ringwald}
\affil{Department of Physics, California State University, Fresno,
	2345 E. San Ramon Ave., MS MH37, Fresno, CA 93740-8031, USA. }
%\email{aastex-help@aas.org}


\altaffiltext{1}{Also Columbia Astrophysics Laboratory, Columbia University,
	550 West 120th Street, New York, New York 10027, USA.}
\altaffiltext{2}{Also Universities Space Research Association}


\begin{abstract}

DQ~Herculis (Nova Herculis 1934) is a deeply eclipsing cataclysmic
variable containing a magnetic white dwarf primary.
The accretion disk is thought to block our line of sight
to the white dwarf at all orbital phases due to its extreme
inclination angle.   Nevertheless, soft X-rays were detected
from \dqher\ with \rosat\ PSPC.  To probe the origin
of these soft X-rays, we have performed \chandra\ ACIS observations.
We confirm that \dqher\ is an X-ray source.  The bulk of the X-rays are
from a point-like source and exhibit a shallow partial eclipse.
We interpret this as due to scattering of the unseen central X-ray source,
probably in an accretion disk wind.  At the same time, we observe
what appear to be weak extended X-ray features around \dqher, which
we interpret as an X-ray emitting knot in the nova shell.

\end{abstract}




\section{Introduction}

A classical nova eruption (see \citealt{S1989} for a review)
is a thermonuclear runaway in a cataclysmic variable (CV;
a semi-detached binary in which a white dwarf primary accretes
from a late-type, usually M dwarf, secondary).  The runaway is
caused when $\sim 10^{-4}$ M$_\odot$ of hydrogen-rich gas is accreted
onto the white dwarf surface; this material becomes degenerate, until
the pressure and temperature at the base of this envelope starts
the runaway reaction.  A nova eruption ejects most of the accreted
envelope.  After a poorly known recurrence time (10$^3$ -- 10$^5$ years),
the underlying binary accumulates enough fresh fuel to erupt again.
Novae are of interest to nuclear astrophysicists; they are useful as
extragalactic distance indicators; they play a role in Galactic
chemical evolution; and they are likely sources of $\gamma$-rays.
To understand a nova, it is essential to study the underlying
system away from its eruption.  Conversely, the study of nova
eruptions is essential to understand the secular evolution of CVs.

The subject of this paper, DQ~Herculis (Nova Herculis 1934),
has played a key historical role in the studies of nova eruptions
and of post-nova binaries (hereafter old novae) due to its
brightness and fortuitous viewing geometry.
The 1934 classical nova eruption of \dqher\ reached $m_v \sim 1.3$
and enabled extensive observations (see \citealt{M1989} for a review).
The discovery of eclipses in \dqher\ \citep{W1956} was
pivotal in establishing the standard picture of CVs.  \cite{W1956}
also discovered the coherent 71 sec photometric oscillation:
we now understand that \dqher\ contains a magnetic white dwarf, accreting
onto the magnetic polar regions which are offset from the rotational poles,
i.e., an oblique rotator.  The ejecta of the 1934 eruption are visible
in H$\alpha$ and other emission lines as an elliptical shell: as of 1978,
43 years after the eruption, the equatorial and polar radii
(expansion velocities) of the \dqher\ shell
were $6.3\pm 0.2$ arcsec ($384\pm 21$ km\,s$^{-1}$)
and $8.6\pm 0.2$ arcsec ($528\pm 25$ km\,s$^{-1}$),
respectively \citep{HS1992}.  They found no evidence for a
deceleration of the expansion, and derived a revised distance estimate
of 561$\pm$19 pc (we use this value throughout this paper).
This makes \dqher\ among the nearest old novae known.
By 1995 September (61 years after eruption), the equatorial and polar
radii have grown to $9.1 \pm 0.8$ arcsec and $12.5 \pm 0.8$ arcsec,
respectively, still consistent with no deceleration, according to
our analysis of an archival {\sl Hubble Space Telescope\/} WFPC2 image.

The eclipsing nature of \dqher\ in principle allows us to place
tight constraints on its system parameters.  The optical eclipse width,
measured from mid-ingress to mid-egress as is customary, is
0.110$\pm$0.003 cycles (as compiled from literature by \citealt{Hea1993}).
Assuming that the source of eclipsed light is centered on the white
dwarf, this can be used to constrain the relationship between the
mass ratio $q = M_2 / M_1$ (where $M_1$ and $M_2$ denote the masses
of the white dwarf and the mass donor, respectively) and the inclination
angle $i$.  \cite{Hea1993}
measured the radial velocity amplitude of the secondary $K_2$
through the detection of the Na\,I infrared doublet absorption
lines.  This, combined with the eclipse width and the $K_1$ from
literature, have led to their estimates of $q = 0.66 \pm 0.04$,
$i = 86^\circ .5 \pm 1^\circ .6$, $M_1 = 0.60 \pm 0.07$ M$_\odot$,
and $M_2 = 0.40 \pm 0.05$ M$_\odot$.

About 20 X-ray bright CVs with coherent spin modulations have been
discovered since the late 1970s, and are classified as intermediate
polars (IPs) or \dqher\  stars \citep{P1994}.  They are among the
brightest CVs in the 2--10 keV band, with luminosities generally
in excess of $10^{32}$ ergs\,s$^{-1}$.  However, the prototype \dqher\ was
{\sl not\/} detected in an \einstein\ Imaging Proportional Counter (IPC)
observation with a 2$\sigma$ upper limit
of 0.0046 ct\,s$^{-1}$ \citep{Cea1981}.  One of their hypotheses,
that high inclination may be the cause of this non-detection, found
support in a model of the phase shifts of the 71-s oscillation during
eclipse \citep{P1980}.  In this model, a beam of high energy photons
originating on the white dwarf is reprocessed by the accretion disk
and results in the optical oscillation.  \cite{P1980} finds a good fit
at $i \sim 89^\circ$, although the actual value of $i$ can be lower,
as long as the entire front half of the disk including the white
dwarf is obscured by the outer edge of the disk.  Such permanent
obscuration of the white dwarf would greatly reduce the X-rays
reaching Earth.

Despite the high inclination angle, X-rays were detected
from \dqher\ with the \rosat\ Position Sensitive Proportional Counter
(PSPC; \citealt{Sea1996})\footnote{\cite{P1994} was already aware
of this detection, citing ``Patterson and Eracleous 1994, in preparation.''}.
The inferred luminosity (0.1--2.0 keV) of
$\sim 4.5 \times 10^{30}$ ergs\,s$^{-1}$ is modest for an IP.
We note that, using a thermal plasma model with
$kT = 0.3$ keV with \nh =1.0 $\times 10^{20}$ cm$^{-2}$,
consistent with the PSPC spectrum, and the revised distance of
561 pc (both are different from the values assumed by \citealt{Cea1981}),
the \einstein\ IPC upper limit corresponds to
2.8 $\times 10^{30}$ ergs\,s$^{-1}$ in the IPC band (0.4--4 keV),
or 3.3 $\times 10^{30}$ ergs\,s$^{-1}$ in the 0.1--2.0 keV band.
These numbers may reflect source variability or the limit of accuracy of
cross-instrument comparisons; neither would be surprising.
A deep eclipse is not observed, consistent with the permanent obscuration
of the white dwarf, suggesting an alternative origin of the observed X-rays.
Although \cite{Sea1996} considered a likely origin to be the secondary,
it has to be emitting X-rays at significantly above the saturation
limit ($10^{-3}$ L$_{bol}$, or a few times 10$^{29}$ erg\,s$^{-1}$ in
this case) seen in rapidly rotating late type dwarfs
(see, e.g., \citealt{Sea1999}).  If not the secondary, what is the origin
of the soft X-rays observed from \dqher?

Only $\sim$150 photons were detected in the \rosat\ observation,
insufficient for further progress.  We have therefore
secured a \chandra\ {X-ray Observatory\/} Advanced CCD Imaging Spectrometer
(ACIS) observation of \dqher.  We describe the observation and data reduction
in \S2, present the results in \S3, and discuss the implications in \S4.

\section{Observations}

We observed \dqher\ with \chandra\ \citep{Wea1996} ACIS without a grating,
placing the object on the back-illuminated S3 chip\footnote{For details
of the ACIS instrument, see the web site at
http://www.astro.psu.edu/xray/acis.}.  The observations
were performed as two pointings, one from 2001 July 26 13:00 UT to
July 27 02:31 UT and the other from 2001 July 29 17:09 UT to
July 30 02:27 UT, for a total integration time of 68.8 ksec.
These observations were performed
at near identical pointing positions and roll angles.
We took the outputs from the processing pipeline at the \chandra\ X-ray
Center, and extracted images, spectra, and light curves
from these observations individually, before combining the results.

We detect a source at RA = 18h07m30.24s, Dec=+45d51m32.6s using
the Ciao routine {\tt celldetect}.  Although this is offset by 1.7$"$
from the position of \dqher\ given in \cite{Dea2001}, this is likely
to reflect the limitation of comparison of  positions measured using
different astrometric reference frames (Downes 2002, private communication).
We identify this X-ray source with \dqher\ based on this positional
coincidence and based on the orbital variability (see \S 3.2 below).

We extracted source spectra and light curves from a 5 arcsec radius
circular region, and background from an annulus of 40 arcsec outer radius
and 20 arcsec inner radius centered on the source.  The size of the
background annulus is chosen to exclude the nova shell emission (see \S 3.3).
The source count rate is 0.025 cts\,s$^{-1}$
(3 $\times 10^{30}$ ergs\,s$^{-1}$
in the 0.2--5 keV band, based on the spectral model discussed below),
while the background in the source region (scaled from the counts in
the background region by the detector areas) is estimated to be
8.2$\times 10^{-4}$ cts\,s$^{-1}$.  \dqher\ did not show a significant
difference in count rate between the two observations.

\section{Results}

\subsection{Spectral Characteristics}

There is a bump in the observed ACIS-S spectrum of \dqher\ around 1 keV
(left panel, Figure 1) which is not in the instrument response.  Because
of this, the spectrum cannot be fitted with smooth continuum models
(power-law, bremsstrahlung or black body) alone.  Using models of
optically thin thermal plasma emission, such as {\tt mekal}
\citep{Mea1985,Mea1986,Lea1995}, the 1 keV bump is interpreted as
a complex of lines in the 0.6--1.2 keV range, notably lines of
O\,VIII (0.64 keV), Fe\,XVII (0.73 \& 0.83 keV),
and Ne\,IX (0.90 keV).  These lines, however,  are weaker relative to
the continuum than the predictions of single-temperature plasma models.
Moreover, the continuum above 2 keV is much stronger than expected for
plasma temperatures that give rise to these lines (kT$\sim$0.6 keV).
As a result, a single component {\tt mekal} model fit also fails to
reproduce the observed spectrum.

The simplest successful representation ($\chi^2_\nu$=1.4; see Figure 1)
is a two-component model\footnote{We have also included the {\tt acisabs}
model in the fit, to account for instrumental absorption due to contaminant,
at the appropriate level.} consisting of kT=0.59$\pm$0.04 keV {\tt mekal}
plasma model plus a power law of photon index 2.7$\pm$0.2, with
\nh\ = 2.9 $\pm 1.2\times 10^{20}$ cm$^{-2}$.  The power-law component
in our model provides the added continuum necessary to achieve an acceptable
fit, both near the lines (0.8--1.2 keV) and at higher energies (2--5 keV).
Inferred intrinsic flux in the 0.2--5 keV band is 8.2 $\times 10^{-14}$
ergs\,cm$^{-2}$s$^{-1}$, or a luminosity of 3.1 $\times 10^{30}$
ergs\,s$^{-1}$.  \dqher\ was not significantly
detected above 5.0 keV (we obtain a count rate and an estimated 1$\sigma$
error in the 5--8 keV band of $0.3 \pm 2.0 \times 10^{-4}$ cts\,s$^{-1}$),
although this is in part due to the lower effective area of the
\chandra\ ACIS instrument at high energies.  We take
$2.0 \times 10^{-4}$ cts\,s$^{-1}$ as our upper limit, which corresponds to
$4 \times 10^{29}$ ergs\,s$^{-1}$ assuming a power law spectral shape with
a photon index of 2 and a source distance of 561 pc.

Note, however, that our adopted model for the spectrum is not unique.
Given the modest number of detected photons ($\sim$1700), and instrumental
limitations (both in bandpass and in spectral resolution), there can be
a wide range of models that fit the data.  We are only confident that
\dqher\ shows a spectral feature around 1 keV which mimics a
collisionally excited, optically-thin plasma of kT$\sim$0.6 keV.
The presence of such a plasma would also lead to a Si line at 1.85 keV,
which may also be present in the data.
Consequently, spectral modeling alone of the \chandra\ data is insufficient
to uncover the origin of the soft X-rays in \dqher.

\subsection{Timing Characteristics}

We show, in Figure 2, the ACIS light curve of \dqher\ folded on the orbital
ephemeris of \cite{Zea1995} into 16 bins per cycle.  The ephemeris has
an  estimated accuracy of $\pm$11 s for this epoch (see \S 4.1 for further
details).  A partial eclipse, lasting for about 3 bins and about 30\% deep,
is clearly seen.   In this representation, the X-ray eclipse center appears
to be offset from the ephemeris prediction by one half of
the bin size, i.e., 0.03 cycle or 500 s.  Moreover, the X-ray eclipse
appears to be wider (3 bins, or 0.1875 cycles) than the optical eclipse
(0.110$\pm$0.003, according to a compilation
in \citealt{Hea1993}), which should be 2 bins in Figure 2.

To quantify this impression, we have fit the 32 bin per cycle version
of the light curve using a piecewise linear function with symmetric
ingress and egress; we show our best-fit model in Figure 2.  The eclipse
center is found to be at phase 0.033$^{+0.030}_{-0.027}$ (90\% confidence),
confirming the offset.  However, the eclipse width (mid-ingress to mid-egress)
is measured to be 0.18$^{+0.05}_{-0.12}$ (90\%) cycle, thus we cannot
claim that the X-ray eclipse is definitely wider than in the optical.
In addition to the formal uncertainties, there is additional uncertainty
arising from our choice of specific model, thus both these numbers need
confirmation.  Our \chandra\ data are of insufficient quality for us to
be able to derive further details of the eclipse profile, such as
ingress/egress durations.

\subsection{Imaging Characteristics}

At first glance, the \chandra\ ACIS image of \dqher\ looks like
that of a point source.  Indeed, over 70\% of the flux is contained within
1 arcsec ($\sim$560 AU, or 8.4$\times 10^{15}$ cm) of the image centroid,
as expected for a soft point source.  However, a closer inspection of
the image shows that there is a region about 2 arcsec by 2 arcsec in
area and about 10 arcsec NE of the point source with more photons (28)
than are expected by chance ($\sim$8).  These excess counts appear equally
distributed over this area, and does not display the sharply peaked core
of the point spread function (PSF) expected of a point source.

We have investigated this further by constructing radial profiles of
the ACIS image of \dqher\ in several energy bands.  We then fit a model
consisting of a flat background plus the PSF available in the \chandra\ 
calibration database (interpolated in energy and off-axis angle as
appropriate).  We find notable excesses in the azimuthally-averaged
radial profiles in the 0.2--0.5 and 0.5--0.8 keV bands (Figure 3;
we find no significant features at energies above 0.8 keV, although
this may in part due to poor counting statistics).  The statistical
significance of this is unfortunately difficult to establish: our
experiences with other bright sources is that PSF fit is usually
less than perfect because of effects of pile-up, background non-uniformity,
uncertainties in the point source location, and imperfect calibration of
the PSF.  However, the structures (clusters of several points above the
best fit models) seen in the radial profiles of \dqher\ is unusual.

The two features marked with horizontal bars (one at 5--9 arcsec in
the 0.2--0.5 keV band, and another at 8--10 arcsec in the 0.5--0.8 keV
band) contain 18.0$\pm$4.6 and 12.2$\pm$4.1 counts, respectively.
Other possible features either have too few photons (the
6--7 arcsec bin in the 0.5--0.8 keV profile has 5) or outside the optically
detected nova shell (12--20 arcsec excess in the 0.2--0.5 keV profile)
and will not be discussed further.  For the two features, we also
construct an azimuthal histogram in 30 degree bins of photons falling
within the appropriate radial and energy bins.  We plot this in the
insets of Figure 3 as a histogram in polar coordinates (the position
angle on the graph corresponds to the position angle on the sky, and
radial distance from the center is proportional to the number of photons
in that bins).  We find that the 0.2--0.5 keV feature is about 8 arcsec
away S -- SW of \dqher, while the 0.5--0.8 keV feature is about 9 arcsec
away NE of the binary.

\section{Discussion}

\subsection{Phase Offset}

The folded X-ray light curve appears to be offset from the ephemeris
prediction by about 0.03 cycles, or 500 s.  The phase offset is greater
than 0.006 cycles, or 100 s, at 90\% confidence level.  Our check of the
data reduction processes (such as the barycentric correction) has not
revealed any potential systematics that could account for such a large
offset.  The linear ephemeris, taken from \cite{Zea1995}, has a nominal
error of 11 s for this epoch.  Moreover, their linear and quadratic
ephemerides have diverged by less than 15 s.  Even though a long-term
sinusoidal modulation in O$-$C eclipse timings has been found
\citep{Pea1978}, the magnitude of this effect is 70 s.
Thus, we believe that the phase offset likely does not result from an
inaccurate ephemeris.

Instead, we consider it likely that the phase shift of the X-ray light
curve is caused by departure from axial symmetry of the \dqher\ system.
We see a considerable scatter in the times of the optical eclipse,
up to 170 s in the O$-$C timings, although some of it
can be modeled by the sinusoidal term (see Figure 4 of \citealt{Zea1995}).
Moreover, the eclipse light curves (see their Figure 1) are obviously skewed,
with ingress beginning around phase $-$0.08 and egress completing
around phase +0.12.  This clearly demonstrates the departure from axisymmetry
of the eclipsed object.  We might still argue that the deepest point in
the eclipse light curve marks the position of the unseen white dwarf.
However, this is not necessarily the case, particularly if the eclipsed
object is the back side of an asymmetric accretion disk \citep{P1980}.
Interestingly, the radial velocity study of the secondary, using the Na\,I
infrared doublet, shows that the inferior conjunction of the secondary is
at phase 0.013$\pm$0.009 on the eclipse ephemeris \citep{Hea1993}.
This possible phase offset between the radial velocity curve of the
secondary and the optical eclipse should be studied further.  In any
case, a non-axisymmetric disk can result in a phase offset between
the optical and X-ray eclipses.

Our best-fit X-ray eclipse width is also larger than the optical eclipse
width, although in this case they can be equal within 90\% confidence
limits.  We do not necessarily expect the two to be equal, since the model
of \cite{P1980} for the phase shift of the 71-s oscillations suggests
that the emission-weighted center of eclipsed light is somewhere on the
back half of the accretion disk, not centered on the white dwarf.
If true, the inclination angle $i$ derived with the latter assumption
\citep{Hea1993} would be an underestimate for a given $q$.  We cannot
solve this difficulty with the data in hand, so we treat this as an extra
source of uncertainty in the analysis to follow.

\subsection{Wind-Scattered X-rays}

At this inclination, a partial eclipse is a signature of an extended emission
region.  Any X-ray emission or scattering from the surface of the accretion
disk should be as deeply eclipsed as the optical light from the disk.
Since the X-ray eclipse is much shallower than in the optical, we
conclude that the X-rays must originate in a vertically extended
region.  Assuming the best-fit system parameters of \cite{Hea1993},
we estimate that materials that are $> 0.28 a$ (where $a$ is the binary
separation) above the orbital plane should remain uneclipsed.  The limit
could be somewhat larger if the inclination is underestimated by
\cite{Hea1993}.

Is this a single, extended X-ray source, or can there be an eclipsed
component and an uneclipsed component?  From the data in hand, we cannot
rule out the latter possibility.  We could test this if we had a higher
quality X-ray light curve of the eclipse. A two-component model would
predict a sharp ingress and egress, while a single extended source would
result in a more gradual eclipse transitions.  Here we proceed assuming
a single extended source, since we do not have a physical model corresponding
to the two-component model.

The partial eclipse of UV emission lines in \dqher\ also demonstrates
the presence of an extended emission region \citep{CM1985, Sea1996,
Eea1998}.  The last authors, in particular, have analyzed the time-resolved
profile of the C\,IV $\lambda$1550 line through the eclipse.  At mid-eclipse,
this line is clearly redshifted, which they argue is a strong indication
that the residual line flux originates in an accretion disk wind (see also
Figure 7. of \citealt{CM1985}), even though an accretion disk wind is usually
seen only in non-magnetic CVs with high accretion rate (see, e.g.,
\citealt{DP2000}).  These systems have an accretion disk that
extends all the way down to the white dwarf surface and an
optically thick boundary layer.  The presence of wind in \dqher,
a system with no boundary layer and a luminous but truncated disk,
suggests that the accretion geometry does not dictate whether
a CV has a wind or not \citep{Eea1998}.  \dqher\ appears to be sufficient
luminous to drive a wind, as the fluxes of the He\,II recombination lines
in \dqher\ requires an ionizing luminosity of order
$10^{34}$ ergs\,s$^{-1}$ \citep{Sea1996, Eea1998}.

The presence of the accretion disk wind in \dqher\ might make this
an unusual IP, since we are not aware of any reports of accretion
disk winds in other IPs.  Instead, it suggests a natural
parallel with the dwarf nova OY~Car in superoutburst and the nova like
system UX~UMa.  Both are bright non-magnetic CV with accretion
disk wind; both are deeply eclipsing in the optical yet
no X-ray eclipses were observed in OY~Car in supreoutburst
\citep{Nea1988, Pea1999} or in UX~UMa \citep{Wea1995}.
This is interpreted to imply that the central X-ray source
in these systems cannot be seen directly, but instead is seen
indirectly after scattering in an extended corona or wind.
The \euve\ observation of OY~Car in superoutburst
\citep{MR2001} confirms this basic picture, since the observed spectrum
can be modeled as resonant scattering of continuum photons in
an accretion disk wind.  A wind, rather than a static corona,
is strongly favored since the EUV emission lines are seen to have
FWHM of $\sim$2300 km\,s$^{-1}$.  We believe the same resonant scattering
model can be applied to the observed soft X-rays from \dqher, except that
the nature of the unseen central source must be different (see \S 4.3).

A static corona can also be the scattering site for the X-ray
photons observed in \dqher.  However, that would make it different
from OY~Car, where the EUV line widths prove the wind to be
responsible for scattering.  Secondly, the UV line profile during
the eclipse is indicative of a wind, while we do not know if a static
corona exists in this system.  Finally, the depth of the partial
eclipse demands a large vertical extent ($> 0.28 a$, see above)
for the scattering region.  This is well away from the disk and
probably is an unlikely location for a static corona.

We therefore interpret the observed soft X-rays as due to
resonant scattering of an unseen central source.  Furthermore,
we prefer the accretion disk wind as the scattering site,
pending final confirmation in the form of X-ray line widths.
It is interesting to note that a partial X-ray eclipse is detected
in \dqher, while none has been found in OY~Car or UX~UMa.  It is not
clear, however, whether this indicates some difference in the wind
geometry compared with between \dqher\ and non-magnetic systems,
or if this is simply due to the more extreme inclination angle
in \dqher.

\subsection{Non-Detection of Reflected X-rays}

The observed X-ray luminosity of 3--5 $\times 10^{30}$ ergs\,s$^{-1}$
is likely to be a small fraction of the intrinsic luminosity of the
central source, although this depends strongly on the unknown efficiency
of scattering in the wind.  The flux of the He~II~$\lambda 1640$ line
has been used to estimate an ionizing luminosity of 1--2
$\times 10^{34}$ ergs\,s$^{-1}$ \citep{Sea1996, Eea1998}.

Can this be in the form of a hard component (kT$\sim$20 keV bremsstrahlung)
usually seen in IPs?  To investigate this question, we have modeled the
reflection of such a component off the surface of the Roche-lobe
filling secondary, applying the method of \cite{Sea2001} developed
for Hercules X-1.  X-rays hitting the cold material on the
surface of the secondary can be scattered or absorbed; since softer
photons are more likely to be absorbed, the scattered component is
harder than the incident component.  We have evaluated the 5--8 keV
flux expected from reflection from the secondary as a function of
the orbital phase, taking into account the shadowing by the accretion
disk and the Roche-lobe filling shape of the secondary, assuming the
system parameters of \cite{Hea1993}, as listed in \S 1.  For the computational
details, see \cite{Sea2001}.

The reflection component should be most prominent at the highest energy
band of \chandra, 5--8 keV.  The viewing geometry for the reflection
component is most favorable at orbital phase 0.5; it would result in
a strong, sinusoidal orbital modulation.  However, no such component
is detected in our \chandra\ data in the 5--8 keV data.  This can be
converted to the upper limit of the bolometric luminosity of the hard
component as a function of the disk thickness (Figure 4).  Taking our
upper limit in the phase bin around phase 0.5, 1/16th cycle wide, to be
7.0$\times 10^{-4}$ cts\,s$^{-1}$, we estimate an upper limit to the
kT=20 keV bremsstrahlung luminosity of 10$^{33}$ ergs\,s$^{-1}$ for
a disk $H/R$ (height over radius) of less than 0.23, or half-opening 
angle $< 13^\circ .0$ ($< 14^\circ .6$ for a kT=50 keV bremsstrahlung
due to the different bolometric correction).   If the intrinsic hard 
component luminosity is 10$^{34}$ ergs\,s$^{-1}$, the reflected luminosity
can be reduced to an undetectable level if the disk half opening angle is
$>17^\circ .7$ for kT=20 keV ($>18^\circ .0$ for kT=50 keV),
almost completely shadowing the secondary.

In comparison, the eclipsing
and dipping low-mass X-ray binary EXO\,0748$-$676 has an inclination
of $i \sim 75^\circ$ \citep{Pea1986}: The presence of X-ray eclipses
require the opening angle of the disk to be less than 15$^\circ$ toward
the secondary in this system.  Although azimuthal structures at such heights
are commonly inferred (including one in U Gem 25$^\circ$ above the disk;
\citealt{Szea1996}), we are not aware of claims for disks thicker than
15$^\circ$ in the direction toward the secondary.  We therefore conclude
that the central hard X-ray component is less luminous than inferred from
the He~II~$\lambda 1640$ line, unless the disk in \dqher\ has an unusually
large thickness.  Further studies of the disk thickness, however, are
required to tighten the upper limit of the hard component luminosity.

In contrast, the lack of reflection component does not allow us to put
a useful constraint on the central luminosity, if it has a soft spectrum.
Accretion luminosity in \dqher\ would be dominated by a soft, blackbody-like
component, if the shock is buried within the white dwarf atmosphere.
In addition, the entire photosphere of the primary may still be hot
from the thermonuclear runaway of 1934. The soft component in AM Her
type magnetic CVs have a typical temperature of 20--30 eV \citep{M1999};
even using a kT=100 eV blackbody, the spectrum is too soft to result in
appreciable reflection signature.  Therefore, our results are consistent
with the ionizing luminosity of 1--2 $\times 10^{34}$ ergs\,s$^{-1}$
\citep{Sea1996, Eea1998} if this is in the form of a soft component,
regardless of the disk height.

\subsection{The Nova Shell around \dqher}

We appear to have detected extended emission features, one about 8 arcsec
S to SW of, and another about 9 NE of \dqher.   The latter location
corresponds to a bright [N\,II] knot in the \dqher\ shell, while the
former has no obvious optical feature associated with it \citep{Sea1995}.
We consider it likely that these small excesses are real and represent
X-ray knots in the \dqher\ shell.  A similar, but much brighter, X-ray
shell has been seen around GK~Per (Nova Persei 1901; \citealt{BO1999}).
This shell has a soft thermal spectrum, with a prominent Ne\,IX line at
0.9 keV \citep{B2002}.  We do not have enough photons for a spectral
analysis for the \dqher\ shell, but it appears much softer than
the GK~Per shell.  This could result from a different speed
class (\dqher\ was a slow nova, while GK~Per was a fast nova),
as well as from a different environment around \dqher\ than around
GK~Per.  The luminosity in the \dqher\ knot appears to be a few times
10$^{28}$ ergs\,s$^{-1}$, subject to a large uncertainty in
count rate to flux conversion, due to the unknown spectral shape.

Given the probable discovery of X-rays from the shell,
and the apparent lack of strong hard X-rays from the central source,
it may be useful to revisit the models of UV and optical
spectra of the shell.  For example, \cite{FT1981} assumed
a hard X-ray emitting magnetic CV with a luminosity in
excess of 10$^{34}$ ergs\,s$^{-1}$ as the source of ionizing
photons.  Our work casts some doubt on such a picture.
\cite{Pea1990}, on the other hand, assumed a 10$^5$
K blackbody source, which is consistent with our X-ray data of
the central binary.  However, how do the shell X-rays fit in?

\cite{B2002} interprets the X-rays from the GK~Per shell
as a miniature supernova remnant.  In this model, the
nova ejecta are plowing into the surrounding interstellar
medium and shock heated into X-ray emitting temperatures.
There certainly is ample kinetic energy in the nova ejecta
to power shell X-ray emission, and the ejecta velocities
are fast enough to result in X-ray emitting temperatures.
The photoionization model of UV and optical lines, on the
other hand, suggest a cold shell.  We regard this not as an
outright contradiction, but as an additional evidence for the
co-existence of hot and cold matter in nova shells.
We have long known of spectra that show both hot and cold
components in nova shells \citep{PG1957, W1992}.
More recent imaging results \citep{Sea1995} and theoretical
work \citep{Lea1997} show the importance of Rayleigh-Taylor
instability.  Nova shells are complex, multi-phased entities,
including, as it now appears, X-ray emitting knots.

\section{Conclusions}

Our \chandra\ observation of the old nova \dqher\ have revealed
a partial X-ray eclipse.  We believe that the most likely explanation
is scattering in the accretion disk wind, with a central X-ray
source which is hidden from our view by the accretion disk at
all orbital phases.  In this model, \dqher\ is similar to
OY~Car in superoutburst and to UX~UMa.  With a higher signal-to-noise
observation, such as we can hope to achieve with \xmm, we should
be able to measure the X-ray eclipse profile in detail, and
check the reality of the phase offset between X-ray and optical
eclipses.  We also detect an apparent weak extended feature, which
we interpret as an X-ray emitting knot in the nova shell.  This would
make \dqher\ the second classical nova, after GK~Per,
with an X-ray emitting shell.

%\acknowledgments
%
%Anybody we need to acknowledge...?

\clearpage

% Figure captions
%---------------------------------------------------------

\begin{figure}
\plotone{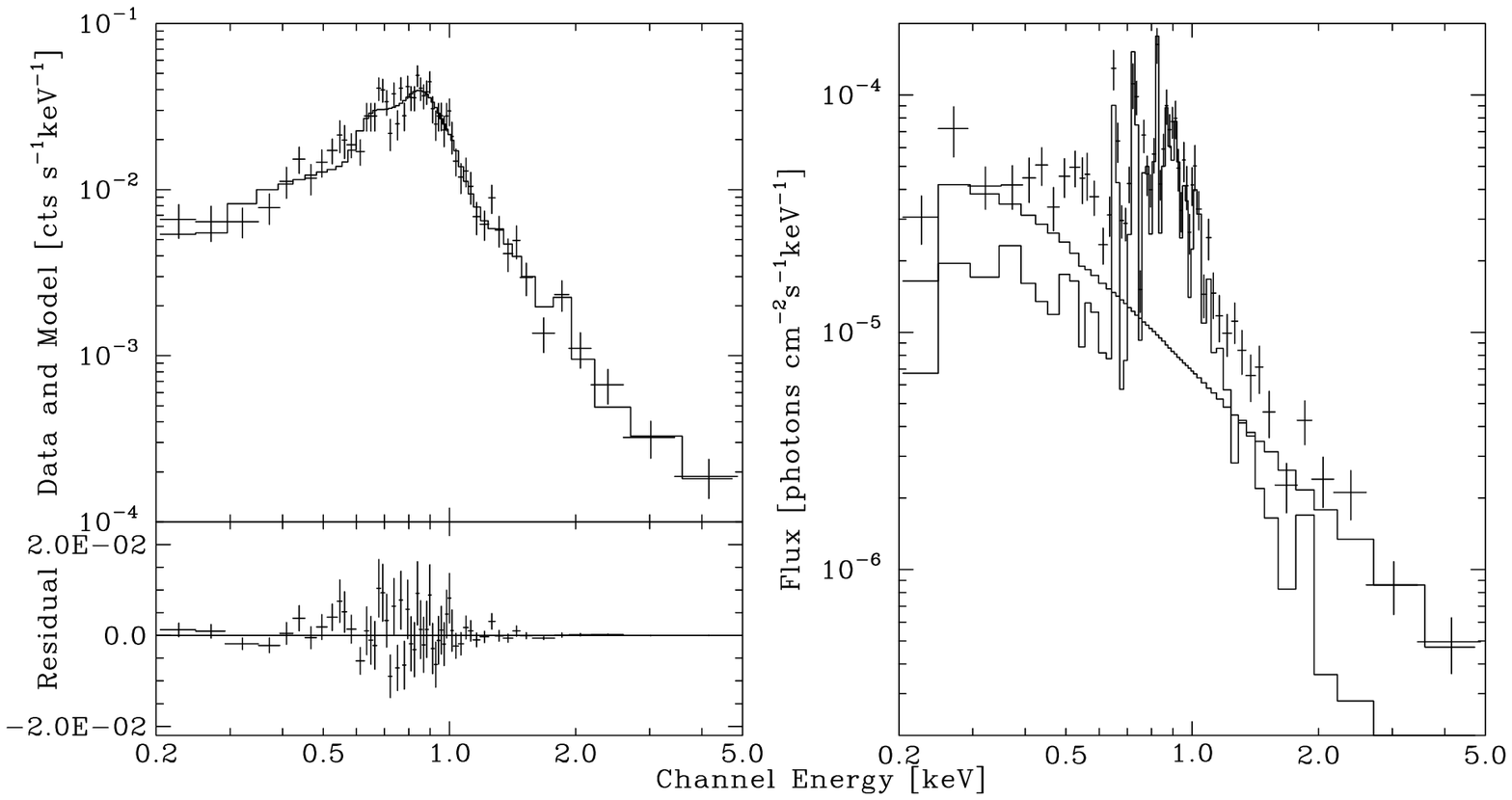}
\caption{The \chandra\ ACIS spectrum of \dqher: (left) The observed
spectrum is shown in upper panel, with the best-fit two component model
folded through the instrument response, while the residuals are shown
in the lower panel; (right) The photon spectrum inferred from the
fit, shown with the power-law and thin thermal plasma components.}
\end{figure}

\begin{figure}
\plotone{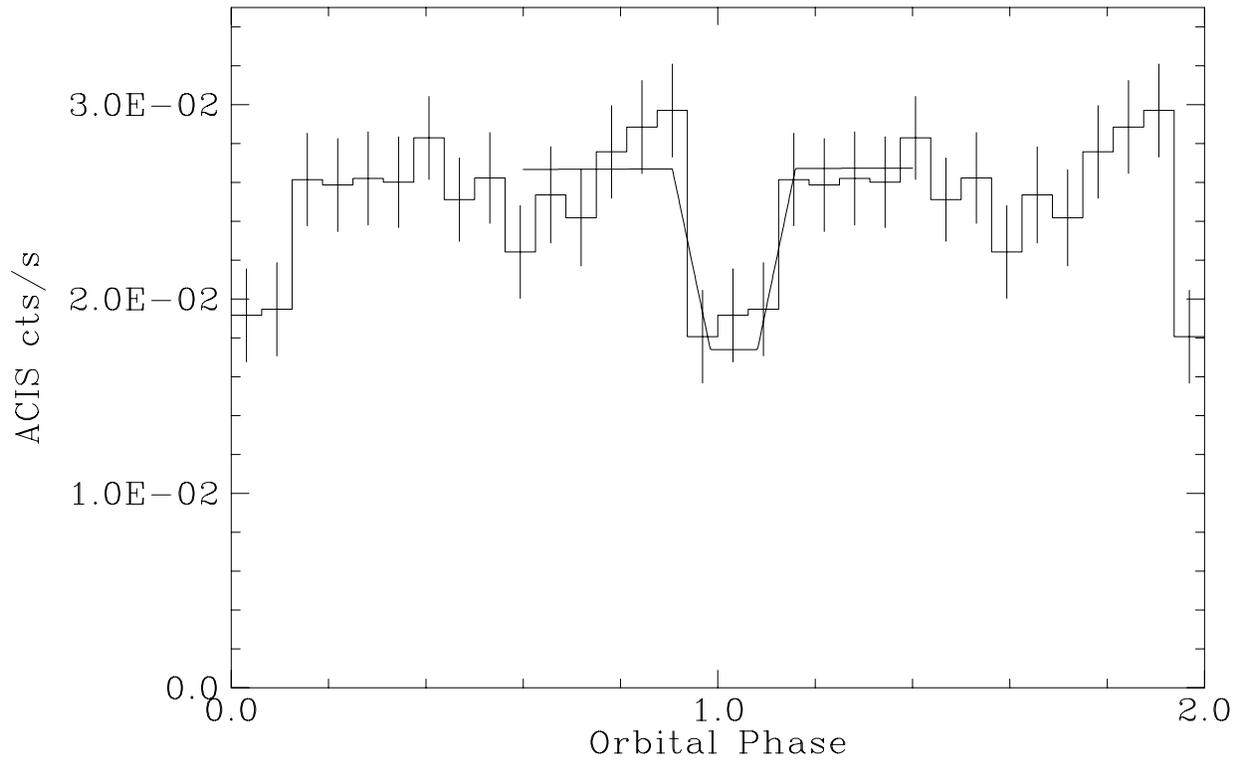}
\caption{The \chandra\ ACIS light curve of \dqher, folded on the
orbital period and plotted twice.  Each bin is 1/16th of the orbital
period.  Also shown is a piecewise-linear model of the eclipse;
the parameters are from the best fit model to the 32 bin per cycle
version of the folded light curve.}
\end{figure}

\begin{figure}
\plotone{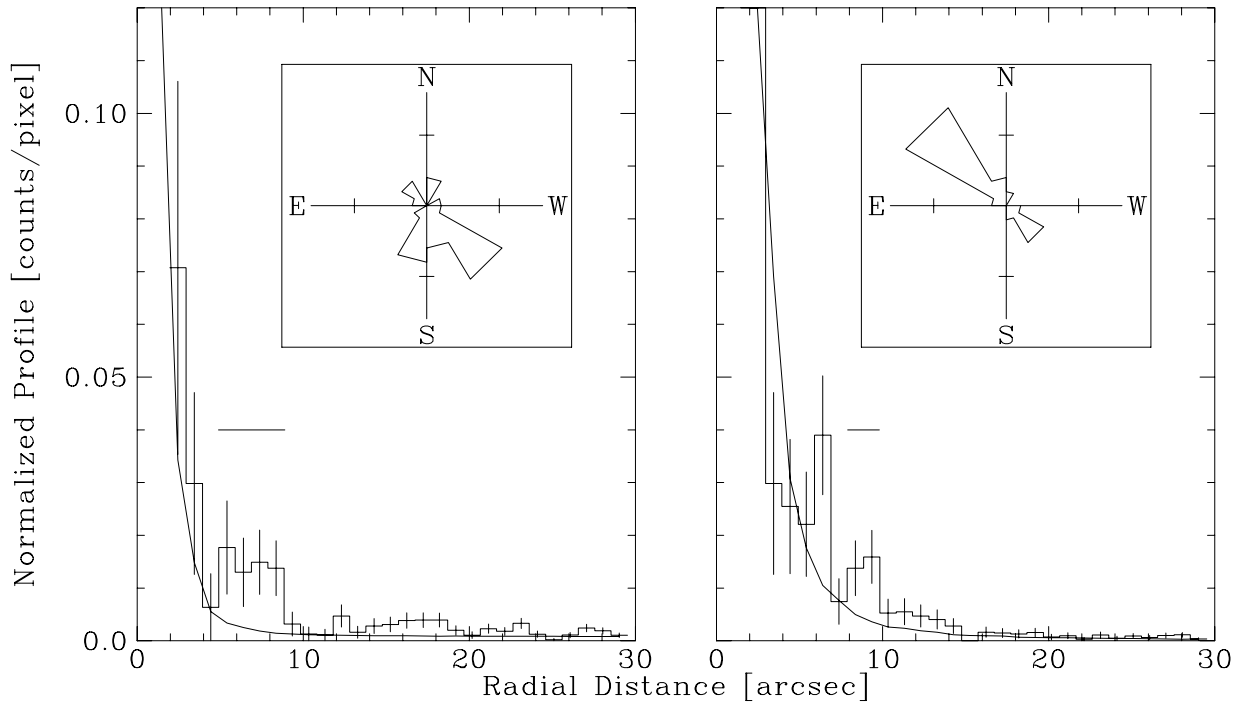}
\caption{The radial profiles of the \chandra\ ACIS-S X-ray image of
\dqher\ in the 0.2--0.5 keV (left) and 0.5--0.8 keV (right) bands.
The data are plotted as a histogram with errors, while the best-fit
model are plotted as a solid line.  The central $\sim$1 arcsec
determines the normalization of the profile, but has been omitted
from this figure to make the wings of the PSF visible.  The horizontal
bars indicate the two prominent features examined further in the insets.
They are histograms in polar coordinates of the excess counts (see text
for details).  Tick marks indicate 5 counts per bin.}
\end{figure}

\begin{figure}
\plotone{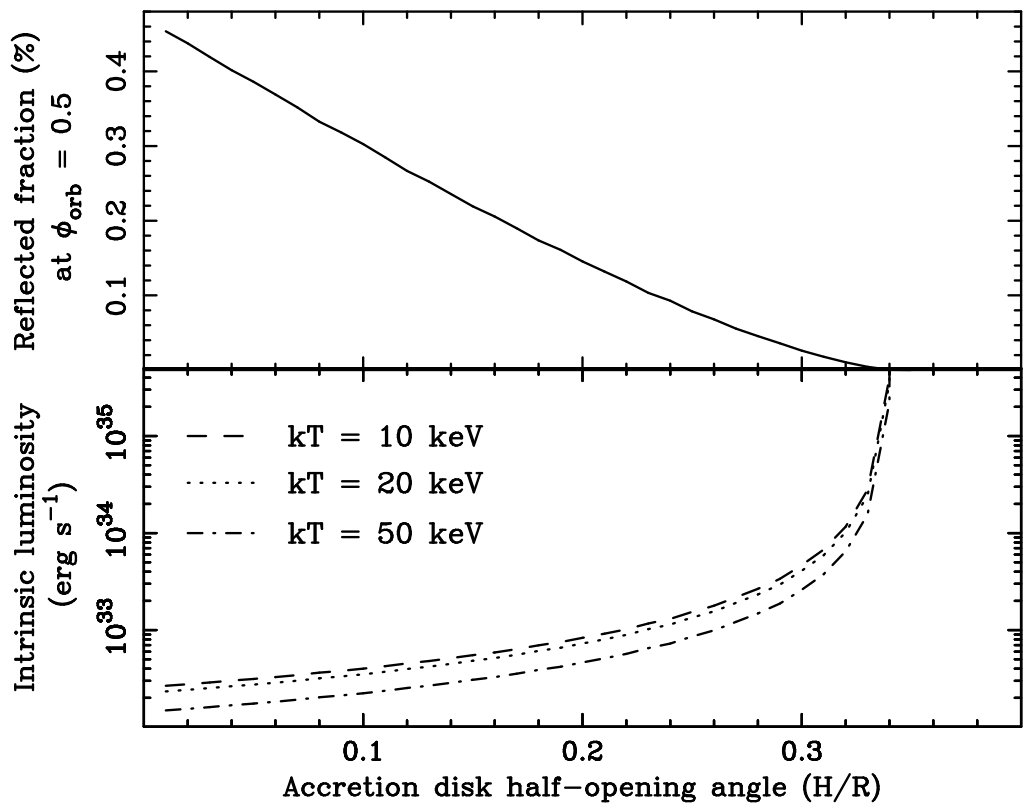}
\caption{(Top) The calculated efficiency of reflection from the secondary
star at orbital phase 0.5.  The reflection fraction in the 5--8 keV band is
plotted as a function of the thickness of the accretion disk. (Bottom):
The inferred upper limit to the central hard component for three different
bremsstrahlung temperatures, as a function of the disk thickness.}
\end{figure}


\clearpage

\begin{thebibliography}

%\bibitem[Bailey (1979)]{B1979} Bailey, J. 1979, \mnras, 187, 645
%% Z Cha photometry -- Eclipse width bible

\bibitem[Balman \& \"Ogelman (1999)]{BO1999} Balman, S. \& \"Ogelman, H.B.
	1999, \apjl, 518, L111
% GK Per Shell with ROSAT HRI

\bibitem[Balman (2002)]{B2002} Balman, S. 2002, in ``High Energy
	Universe at Sharp Focus,'' eds. E.M. Schlegel \& S.D. Vrtilek,
	ASP Conf. Ser., 262, 34
% GK Per Shell with Chandra

\bibitem[C\'ordova \& Mason (1985)]{CM1985} C\'ordova, F.A. \& Mason, K.O.
	1985, \apj, 290, 671
% Accretion Disk Wind

\bibitem[C\'ordova et al.\ (1981)]{Cea1981} C\'ordova, F.A., Mason, K.O.
	\& Nelson, J.E. 1981, \apj, 245, 609
% Einstein non-detection

\bibitem[Downes et al.\ (2001)]{Dea2001} Downes, R.A., Webbink, R.F.,
	Shara, M.M., Ritter, H., Kolb, U. \& Duerbeck, H.W. 2001,
	\pasp, 113, 764
% CV Catalog Living Edition

\bibitem[Drew \& Proga (2000)]{DP2000} Drew, J.E. \& Proga, D. 2000,
	New Astron. Rev. 44, 21
% Wind review

\bibitem[Eracleous et al.\ (1998)]{Eea1998} Eracleous, M., Livio, M.,
	Williams, R.E., Horne, K., Patterson, J., Martell, P.
	\& Korista, K.T. 1998, in ``Wild Stars in the Old West,''
	eds. S. Howell, E. Kuulkers \& C. Woodward, ASP Conf. Ser.,
	137, 438
% HST Spectrophotometry

\bibitem[Ferland \& Truran (1981)]{FT1981} Ferland, G.J. \& Truran, J.W.
	1981, \apj, 244, 102
% Shell model with Hard X-rayys

% \bibitem[Gallagher (1995)]{G1995} Gallagher, J.H. 1995, in Cataclysmic
%	Variables, edited by A. Bianchini, M. Della Valle, and M. Orio
%	(Dordrecht: Kluwer), p. 285
% Can add Gallagher ref for hot & cold matter in nova shell, but
% that's in a conference proceedings and we already have 2 others.

\bibitem[Herbig \& Smak (1992)]{HS1992} Herbig, G.H., \& Smak, J.I. 1992,
	Acta Astron., 42, 17
% Expansion parallax

\bibitem[Horne et al.\ (1993)]{Hea1993} Horne, K., Welsh, W.F. \& Wade, R.A.
	1993, \apj, 410, 357
% Eclipse/mass study

\bibitem[Liedahl et al.\ (1995)]{Lea1995} Liedahl, D.A., Osterheld, A.L.
	\& Goldstein, W.H. 1995, \apjl, 438, 115
% L in mekal

\bibitem[Lloyd et al.\ (1997)]{Lea1997} Lloyd, H.M., O'Brien, T.J.
	\& Bode, M.F. 1997, \mnras, 286, 137
% Theory of Rayleigh-Taylor instability in nova shell

\bibitem[Martin (1989)]{M1989} Martin, P.G. 1989, Classical Novae,
	M.F. Bode \& A. Evans, J. Wiley \& Sons, 1989, p.~93
% DQ Her chapter in the Nova book.

\bibitem[Mauche (1999)]{M1999} Mauche, C.W. 1999, in ``Annapolis Workshop
	on Magnetic Cataclysmic Variables,'' eds. by C. Hellier \& K. Mukai,
	ASP Conf. Ser., 157, 157
% Soft component temperature

\bibitem[Mauche \& Raymond (2001)]{MR2001} Mauche, C.W. \& Raymond, J.C.
	2001, \apj, 541, 924
% OY Car in superoutburst

\bibitem[Mewe et al.\ (1985)]{Mea1985} Mewe, R., Gronenschild, E.H.B.M.
	\& van den Oord, G.H.J. 1985, \aaps, 62, 197

\bibitem[Mewe et al.\ (1986)]{Mea1986} Mewe, R., Lemen, J.R.
	\& van den Oord, G.H.J. 1986, \aaps, 65, 511.
% M in mekal

\bibitem[Naylor et al.\ (1988)]{Nea1988} Naylor, T., Bath, G.T., Charles, P.A.,
	Hassall, B.J.M., Sonneborn, G., van der Woerd, H. \& van Paradijs, J.
	1988, \mnras, 231, 237
% OY Car in Superoutburst

\bibitem[Parmar et al.\ (1986)]{Pea1986} Parmar, A.N., White, N.E.,
	Giommi, P. \&  Gottwald, M. 1986, \apj, 308, 199
% EXO0748 --- disk thickness in LMXBs

\bibitem[Patterson (1994)]{P1994} Patterson, J. 1994, \pasp, 106, 209
% Review

\bibitem[Patterson et al.\ (1978)]{Pea1978} Patterson, J., Robinson, E.L.
	\& Nather, R.E. 1978, \apj, 224, 570
% Sinuosoidal eclipse timing changes

\bibitem[Payne-Gaposchkin (1957)]{PG1957} Payne-Gaposchkin, C. 1957,
	The Galactic Novae, North Holland, Amsterdam
% The real nova bible

\bibitem[Petitjean et al.\ (1990)]{Pea1990} Petitjean, P., Boisson, C.
	\& P\'equignot, D. 1990, \aap, 240, 433
% Nova shell model with soft X-rays

\bibitem[Petterson (1980)]{P1980} Petterson, J.A. 1980, \apj, 241, 247
% Jacobus Peterson probably was the first to propose that the white
% dwarf is forever hidden.

\bibitem[Pratt et al.\ (1999)]{Pea1999} Pratt, G.W., Hassall, B.J M.,
	Naylor, T., Wood, J.H. \& Patterson, J. 1999, \mnras, 309, 847
% OY Car in Superoutburst with ROSAT

\bibitem[Shara (1989)]{S1989} Shara, M. 1989, \pasp, 101, 5
% Classical Nova review

\bibitem[Silber et al.\ (1996)]{Sea1996} Silber, A.D., Anderson, S.F.,
	Margon, B. \& Downes, R.A. 1996, \apj, 462, 428
% UV study with ROSAT detection

\bibitem[Singh et al.\ (1999)]{Sea1999} Singh, K.P, Drake, S.A., Gotthelf, E.V.
	\& White, N.E. 1999, \apj, 512, 874
% Single star saturation limit

\bibitem[Slavin et al.\ (1995)]{Sea1995} Slavin, A.J., O'Brien, T.J.
	\& Dunlop, J.S. 1995, \mnras, 276, 353
% Nova Shell

\bibitem[Still et al.\ (2001)]{Sea2001} Still, M., O'Brien, K., Horne, K.,
	Boroson, B., Titarchuk, L.V., Engle, K., Vrtilek, S.D.,
	Quaintrell, H. \& Fieldler, H. 2001, \apj, 554, 352
% Her X-1 reflected X-rays

\bibitem[Szkody et al.\ (1996)]{Szea1996} Szkody, P., Long, K.S., Sion, E.M.
	\& Raymond, J.C. 1996, \apj, 469, 834
% U Gem dip

\bibitem[Walker (1956)]{W1956} Walker, M.F. 1956, \apj, 123, 68
% Eclipse and pulsation

\bibitem[Weisskopf et al.\ (1996)]{Wea1996} Weisskopf, M. C., O'Dell, S. L.,
	\& van Speybroeck, L. P. 1996, Proc. SPIE, 2805, 2 
% Chandra reference

\bibitem[Williams (1992)]{W1992} Williams, R.E. 1992, \aj, 104, 725
% Hot and cold in nova shell

\bibitem[Wood et al.\ (1995)]{Wea1995} Wood, J.H., Naylor, T. \& Marsh, T.R.
	1995, \mnras, 274, 31
% UX UMa ROSAT

\bibitem[Zhang et al.\ (1995)]{Zea1995} Zhang, E., Robinson, E.L.,
	Stiening, R.F. \& Horne, K. 1995, \apj, 454, 447
% 142 sec? Ephemeris

\end{thebibliography}
\end{document}